\documentclass[paper]{ite}                  
\usepackage{graphicx}
\usepackage[usenames]{color}
\usepackage{amsmath,bm,amsfonts,amssymb}
\usepackage{setspace}

\def\ColorResponseAA{black}
\def\ColorResponseAB{black}
\def\ColorResponseAC{black}

\title{Few-shot Personalized Saliency Prediction\\ Based on \textcolor{\ColorResponseAB}{Interpersonal} Gaze Patterns}
\authorlist{%
 \authorentry{Yuya Moroto}{n}{Company1}
 \authorentry{Keisuke Maeda}{n}{Company3}
 \authorentry{Takahiro Ogawa}{m}{Company3}
 \authorentry{Miki Haseyama}{m}{Company3}
}
\affiliate[Company1]
 {Graduate School of Information Science and Technology, Hokkaido University}
 {Sapporo, Japan}
\affiliate[Company3]
 {Faculty of Information Science and Technology, Hokkaido University}
 {Sapporo, Japan}

\acceptancePreIN{} 

\videoFlag{0} 

\begin{document}

\begin{abstract}
This study proposes a few-shot personalized saliency prediction method that leverages \textcolor{\ColorResponseAB}{interpersonal} gaze patterns. Unlike general saliency maps, personalized saliency maps (PSMs) capture individual visual attention and provide insights into individual visual preferences. However, predicting PSMs is challenging because of the complexity of gaze patterns and the difficulty of collecting extensive eye-tracking data from individuals.
An effective strategy for predicting PSMs from limited data is the use of eye-tracking data from other persons. To efficiently handle the PSMs of other persons, this study focuses on the selection of images to acquire eye-tracking data and the preservation of the structural information of PSMs. In the proposed method, these images are selected such that they bring more diverse gaze patterns to persons, and structural information is preserved using tensor-based regression. The experimental results demonstrate that these two factors are beneficial for few-shot PSM prediction.
\end{abstract}


\maketitle

\section{Introduction}
\label{intro}
Humans can selectively obtain vital information from the abundant visual information in the complex real world through their visual system. Many researchers have attempted to introduce such human mechanisms into image processing models~\cite{itti1998model}~\cite{ullah2020brief}~\cite{borji2019salient}. Specifically, a saliency map, which represents the salient parts that are more noticeable than the neighboring parts, is predicted to reproduce the human instinctive visual perception~\cite{itti1998model}~\cite{harel2007graph}~\cite{hou2012image}~\cite{pan2017salgan}~\cite{kroner2020contextual}. A saliency map is predicted for each image without personalization. However, different individuals focus on different areas even when viewing the same scene, that is, individual differences exist~\cite{li2018personalized}~\cite{risko2012curious}~\cite{le2017visual}. To model individual visual attention, saliency maps have been personalized over the past few years~\cite{xu2017beyond}~\cite{lin2018s}~\cite{moroto2021few}~\cite{Chen_2023_CVPR}~\cite{wang2023predicting}. A traditional saliency map and its personalization are distinguished by referring to them as a universal saliency map (USM) and a personalized saliency map (PSM), respectively. A USM omits individual differences, whereas a PSM is predicted for each person. Personalized visual preferences can be reflected by differences between PSMs~\cite{xu2018personalized}~\cite{gygli2013interestingness}~\cite{li2017towards}; thus, individuality can be useful in many situations. \textcolor{\ColorResponseAB}{Such individuality has significant potential in applications where personal visual preferences play a crucial role. For instance, PSM prediction can be applied to targeted advertising, where understanding individual visual preference helps determine advertisement placements or make UI designs~\cite{10.1145/3379337.3415825,Zheng_2018_ECCV}. Another example is a recommender system, which highlights the regions or contents related to users' interests in images or products~\cite{cho2002personalized,jelassi2013personalized}. Moreover, PSMs can contribute to transferring skills related to tacit knowledge, which is knowledge that is difficult to verbalize but manifests in habitual behaviors or internalized expertise~\cite{nakamura2019pottery,yu2022towards}. These applications demonstrate the practical relevance of predicting PSMs beyond USMs.} Here, to obtain a PSM for unseen images in advance, the PSM should be predicted from the individual gaze pattern tendency.
\par To model individual gaze patterns, the relationship between visual stimuli, e.g., images, and the individual PSM should be analyzed based on eye-tracking data obtained from each person in the past. Then, the gaze patterns emerging in the images are complex and different, and these characteristics lead to the difficulty of PSM prediction. To extract individual gaze pattern tendencies, several researchers have collected eye-tracking data for thousands of images~\cite{xu2017beyond}~\cite{lin2018s}~\cite{Chen_2023_CVPR}~\cite{xu2018personalized}. 
The prediction models adopted in these studies are based on deep learning, which requires a massive amount of training data for each person. The large-scale PSM dataset is openly available; however, the acquisition of a massive amount of individual eye-tracking data can be a significant burden and time-consuming task for new persons in the application. Consequently, a PSM prediction method with a limited amount of training eye-tracking data is required.
\par To predict a PSM from a limited amount of data, an effective strategy is to use PSMs obtained from persons with similar gaze patterns to the target person. To determine whether a person has gaze patterns similar to those of the target person, several pairs of eye-tracking data for the same images are required. However, such pairs cannot be acquired in large quantities, and the selection of images to acquire eye-tracking data is an important process. In a previous study~\cite{moroto2020few}, images that induce the scattering of gazes were selected using adaptive image selection (AIS) to efficiently and steadily obtain the similarity of gaze patterns between the target and other persons (called training persons in this study).  
Additionally, in a previous study~\cite{moroto2021few}, the collaborative multi-output Gaussian process regression (CoMOGP)~\cite{nguyen2014collaborative} was used with the PSMs obtained from training persons for PSM prediction. However, such regression-based methods require vector-format input, and the structural information of PSMs cannot be effectively used.
\par 
Structural information is an important clue to detecting salient areas in the human visual system~\cite{itti1998model}. \textcolor{\ColorResponseAA}{
The method proposed in the previous work~\cite{itti1998model} extracts hand-crafted image features and takes the center-surround differences of them. In this method, as a result of feature extraction, several feature maps are calculated with considering the pixel positions and their relationships, that is, the two-dimensional spatial configuration is preserved. Inspired by this process, we hypothesized that such two-dimensional spatial configuration is useful for PSM prediction and captures the relative positioning and distribution of salient regions within an image. In this way, this paper focuses on the two-dimensional spatial configuration as the structural information for performance improvement of PSM prediction.} Then, it is necessary to construct a PSM prediction method that considers structural information that is compatible with the effective use of PSMs predicted for several training persons. 
Therefore, to improve few-shot PSM prediction, it is desirable to collaboratively incorporate the adaptive selection of images to acquire eye-tracking data and preservation of the structural information of PSMs predicted for training persons.
\par We propose a few-shot personalized saliency prediction method based on \textcolor{\ColorResponseAB}{interpersonal} gaze patterns. In the proposed method, we collaboratively use AIS~\cite{moroto2020few} and the tensor-based regression model~\cite{lock2018tensor}. The AIS scheme focuses on the variety of selected images and the variation in PSMs obtained from the training persons for selecting images. Through the AIS scheme, we can efficiently and steadily obtain the similarity of gaze patterns between the target and training persons. In addition, the input and output of the tensor-based regression model~\cite{lock2018tensor} are in a multi-array tensor format; thus, it predicts the PSMs of the target person from the PSMs of training persons while preserving the structural information. Therefore, we realize the effective selection of images to acquire eye-tracking data and preservation of the structural information of PSMs predicted for training persons. 
\section{Related Works}
\label{related}
\subsection{USM Prediction}
In the field of image processing, USM prediction is a traditional research subject. Specifically, early USM models were constructed based on hand-crafted image features until the development of deep learning methods~\cite{itti1998model}~\cite{harel2007graph}~\cite{hou2012image}. In contrast, deep learning methods, e.g., convolutional neural networks (CNNs), generative adversarial networks (GANs), and vision transformers, have outperformed these models, which do not require training phase~\cite{pan2017salgan}~\cite{kroner2020contextual}. Although many USM prediction methods have been proposed, they have limitations in terms of the performance improvement of PSM prediction because they do not account for individual differences.
\begin{table*}[t]
    \centering
    \caption{\small \textcolor{\ColorResponseAA}{ Comparison of representative methods for USM and PSM prediction.}}
    \small
    \begin{tabular}{lccc}
        \hline
        \textbf{Method} & 
        \textbf{Personalization} & 
        \begin{tabular}{c}
            \textbf{Order of} \\
            \textbf{Training data}
        \end{tabular} &
        \begin{tabular}{c}
            \textbf{Additional}\\
            \textbf{Data}
        \end{tabular} \\
        \hline
        \textbf{USM Prediction} \\ 
        \ \ Computational models~\cite{itti1998model,harel2007graph,hou2012image}  & - & - & - \\
        \ \ Deep learning-based USM prediction~\cite{pan2017salgan,kroner2020contextual} & - & $10^4$ & - \\
        \hline
        \textbf{PSM Prediction} \\ 
        \ \ Multi-task CNN~\cite{xu2017beyond} & \checkmark & $10^3$ / Person & - \\
        \ \ CNN-PIEF~\cite{xu2018personalized} & \checkmark & $10^3$ / Person & Person-specific information \\
        \ \ Sherkati et al.~\cite{sherkati2022clustered} & \checkmark & $10^3$ / Person & Person-specific information \\
        \ \ Strohm et al.~\cite{strohm2024learning} & \checkmark & $10^3$ / Person & Pretrained person embedding network \\
        
        \begin{tabular}{l}
            Person similarity-based approach~\cite{moroto2021few,moroto2020few,moroto2022few} \\(Our setting)
        \end{tabular} & \checkmark & $\bm{10^2}$ / Person & Other person's eye tracking data \\
        \hline
    \end{tabular}
    \label{data_comparison}
\end{table*}
\subsection{PSM Prediction}
\label{psm_prediction_related_works}
The advancement of measurement instruments has sparked interest in PSM prediction over the last decade. The open large-scale dataset significantly contributes to the construction of PSM prediction models~\cite{xu2017beyond}. Specifically, a multi-task CNN-based model achieved highly accurate PSM prediction by focusing on the difference between USMs and PSMs~\cite{xu2017beyond}. In addition, a CNN model with person-specific information encoded filters (CNN-PIEF) was proposed as an extended version of the previous study~\cite{xu2018personalized}. In CNN-PIEF, the embeddings of person-specific information enable the personalization of the prediction model. In addition, in a previous work~\cite{sherkati2022clustered}, a model based on conditional GANs with person clusters constructed from person-specific information was proposed, and in another study~\cite{strohm2024learning}, a siamese CNN-based model for learning user embedding was proposed. These models successfully predicted PSMs using deep learning. However, deep learning requires a sufficient amount of training eye-tracking data to train CNNs, and significant amounts of training data are required to make predictions for a new person. 
\par In this regard, several studies have attempted to reduce the amount of training data by using eye-tracking data obtained from persons with similar gaze patterns to the target person~\cite{moroto2021few}~\cite{moroto2020few}~\cite{moroto2022few}. However, in these methods, the structural information of PSMs cannot be effectively used. Structural information is an important clue to detecting salient areas in the human visual system~\cite{itti1998model}. Thus, it is necessary to construct a PSM prediction method that considers structural information that is compatible with the effective use of PSMs predicted for several training persons. 
\par 
\textcolor{\ColorResponseAA}{
We show the comparison of the representative methods for USM and PSM prediction in Table~\ref{data_comparison}. As shown in the table, while the USM prediction methods can be trained on a large amount of training data by collecting gaze data from various people, the PSM prediction methods should be trained on a limited amount of training data for each person due to personalized prediction. Furthermore, deep learning-based PSM prediction methods~\cite{xu2017beyond,xu2018personalized,sherkati2022clustered,strohm2024learning} still require a large amount of training data, about 1,000 eye-tracking data per person, which is a heavy burden on the collection of training data from new persons. With regard to CNN-PIEF~\cite{xu2018personalized} and Sherkati et al.~\cite{sherkati2022clustered}, PSMs can be predicted by using the person-specific information, e.g., age, and gender, but such information is not always available in the real-world applications. In the study by Strohm et al.~\cite{strohm2024learning}, the person embedding network is pretrained on a large amount of training data. On the other hand, our setting requires a moderate amount of training data, 100 training data per person, but offers  an adequate balance between personalization and scalability. Therefore, the performance comparison is difficult due to the difference of the training data and the research objective.}
\begin{figure*}[t]
\centering
 \includegraphics[clip,width=\textwidth]{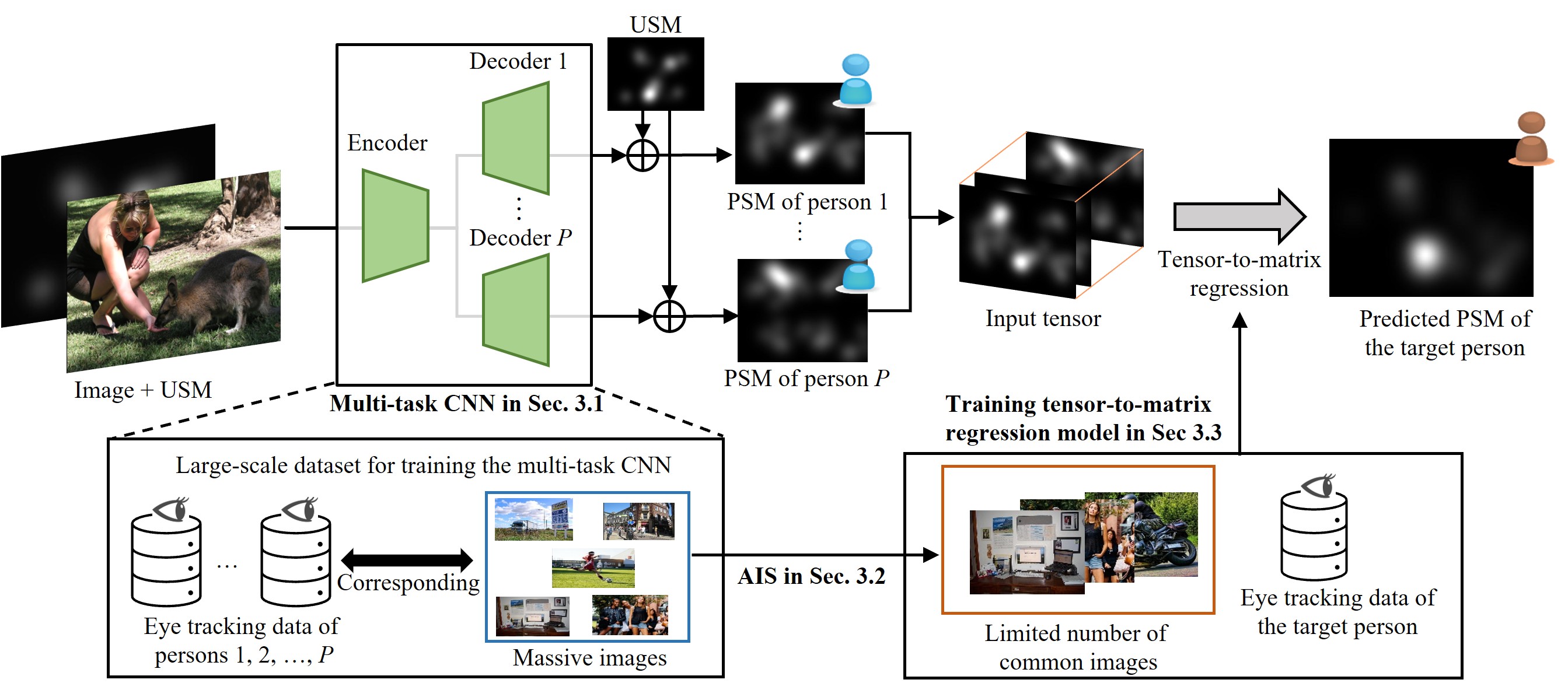}
 \caption{\small Entire flow of the proposed PSM prediction method consisting of three phases. In the first phase, the multi-task CNN~\cite{yin2017multi} predicts the PSMs of $P$ training persons. Next, using the AIS scheme~\cite{moroto2020few}, $I$ images are selected as common images that the target person gazes at. Finally, the PSM is predicted using tensor-to-matrix regression ~\cite{lock2018tensor} with the PSMs of training persons.}
 \label{overview}
\end{figure*}
\section{Proposed Few-shot PSM Prediction}
\label{prop}
The proposed few-shot PSM prediction comprises three phases, and the entire flow is depicted in Fig.~\ref{overview}. Here, we assume that there are $P$ training persons with a massive amount of eye-tracking data and a target person with a limited amount of eye-tracking data.
First, the multi-task CNN~\cite{yin2017multi} is trained to predict the PSMs of the training persons by referring to the previous studies~\cite{xu2017beyond}~\cite{xu2018personalized}.
Next, we select common images that the target person gazes at based on the AIS  scheme~\cite{moroto2020few}. The common images are selected such that they bring more diverse gaze patterns to persons. Finally, the proposed method predicts the PSM using tensor-to-matrix  regression~\cite{lock2018tensor} with the PSMs of the training persons. 
Therefore, by efficiently using the \textcolor{\ColorResponseAB}{interpersonal} gaze patterns, we can effectively select images to acquire eye-tracking data and preserve the structural information of PSMs predicted for training persons. 
\subsection{Multi-Task CNN for Training Persons}
\label{multi-task-CNN}
To train the multi-task CNN model~\cite{yin2017multi},
we prepare the training images $\bm{X}_n \in \mathbb{R}^{d_1 \times d_2 \times d_3}$ ($n=1,2,\dots,N$; $N$ being the number of training images) and their USMs $\bm{U}(\bm{X}_n)\in \mathbb{R}^{d_1 \times d_2}$, where $d_1 \times d_2$ and $d_3$ denote the size of the image and the color channel, respectively. To effectively obtain the predicted PSMs of training persons, previous studies~\cite{xu2017beyond}~\cite{xu2018personalized} have adopted the specific approach of predicting the difference map $\bm{M}(\bm{X})_p \in \mathbb{R}^{d_1 \times d_2}$ ($p=1,2,\dots,P$) between USMs and PSMs as follows:
\begin{align}
    \bm{M}(\bm{X})_p = \bm{S} (\bm{X})_p - \bm{U}(\bm{X}),
\end{align}
where $\bm{S} (\bm{X})_p$ denotes the PSM of the $p$th training person based on the eye-tracking data for the image $\bm{X}$. Next, to simultaneously predict the PSMs of training persons, we construct a multi-task CNN consisting of one image encoder and $P$ PSM decoders and optimize their trainable parameters by minimizing the following objective function:
\begin{align}
    \sum_{p=1}^{P}\sum_{n=1}^{N}\sum_{l=1}^{L}||\hat{\bm{M}}_l(\bm{X}_n)_p-\bm{M}(\bm{X}_n)_p||^2_F,
\end{align}
where $\hat{\bm{M}}_l(\bm{X}_n)_p$ ($l=1,2,\dots,L$; $L$ being the number of convolution layers in one decoder) denotes a predicted difference map calculated from the $l$th layer, and $||\cdot||_F^2$ denotes the Frobenius norm. 
\par Given the test image $\bm{X}_{\textrm{tst}}$, the predicted PSM of the $p$th person is calculated as follows: 
\begin{align}
    \hat{\bm{S}}(\bm{X}_{\textrm{tst}})_p = \hat{\bm{M}}_L(\bm{X}_{\textrm{tst}})_p + \bm{U}(\bm{X}_{\textrm{tst}}). \label{eq2}
\end{align}
Therefore, the multi-task CNN can simultaneously predict the PSMs of the training persons and consider the relationship between these PSMs.
\subsection{Adaptive Image Selection for PSM Prediction}
\label{AIS}
We select a few images from the $N$ training images to obtain the tendency of the target and training persons to be similar. To effectively analyze such similarity, the $I$ common images that produce more diverse gaze patterns to persons are selected using the AIS scheme~\cite{moroto2020few}. Specifically, the AIS scheme focuses on various common images and variations in PSMs obtained from the training persons. To simultaneously consider these factors, the AIS scheme uses the variation in PSMs for objects in each image. 
\par First, we calculate the PSMs and their variance for each object $\bm{B}_{n,j}$ ($j=1,2,\dots,J$; $J$ being the number of object categories in the training images) in the training images $\bm{X}_n$. 
Then, object detection ~\cite{ge2021yolox} is applied to the training images to obtain a rectangle with dimensions of $d^h_{n,j}\times d^w_{n,j}$ for the $j$th object in the $i$th image. The PSM variance $q_{n,j}$ for object $\bm{B}_{n,j}$ is calculated as follows:
\begin{align}
    &q_{n,j} = \frac{1}{d^h_{n,j}d^w_{n,j}P}\sum_{p=1}^P ||\bar{\bm{S}} (\bm{B}_{n,j})_p\odot\bar{\bm{S}} (\bm{B}_{n,j})_p||_F^1, \\
    &\bar{\bm{S}} (\bm{B}_{n,j})_p = \bm{S} (\bm{B}_{n,j})_p-\frac{1}{P}\sum_{p=1}^P \bm{S} (\bm{B}_{n,j})_p,
\end{align}
where $\bm{S}(\bm{B}_{n,j})_p$ denotes the PSM for object $\bm{B}_{n,j}$ of person $p$, and $\odot$ denotes the operator of the Hadamard product. Then, we set $q_{n,j}=0$ when $\bm{X}_n$ does not include the $j$th object and set the largest $q_{n,j}$ when the image $\bm{X}_n$ includes several $m$th objects. 
Then, we obtain the sum of $q_{n,j}$ for $n$th image as follows:
\begin{align}
    \bar{q}_n = \sum_{j=1}^{J} q_{n,j}.
\end{align}
Finally, using $\bar{q}_n$, we select the top $I$ images as common images under the constraint to maximize the number of object categories in these images.
Consequently, the selected common images have multiple object categories, and the objects in these images exhibit a high PSM variance. 
Specifically, the AIS scheme focuses on the PSM variance to analyze the differences in gaze patterns and persons’ visual preferences, and a higher number of object categories leads to greater scene diversity. For images, the AIS scheme has demonstrated its strength when image selection for training images exhibits high diversity. In addition, the AIS scheme is based on object-level gaze analysis, and it contributes to maintaining the diversity of visual information. In contrast, for PSMs, the target person’s gaze pattern can be represented by combining the gaze patterns of multiple persons even if no person has a similar gaze pattern, and this possibility increases with the number of training persons. In addition, if the eye-tracking data of a single person contain noise or outliers, the effect can be reduced in the AIS scheme by focusing on the PSM variance of several persons.
\begin{figure*}[t]
    \centering
    \includegraphics[clip,width=0.9\textwidth]{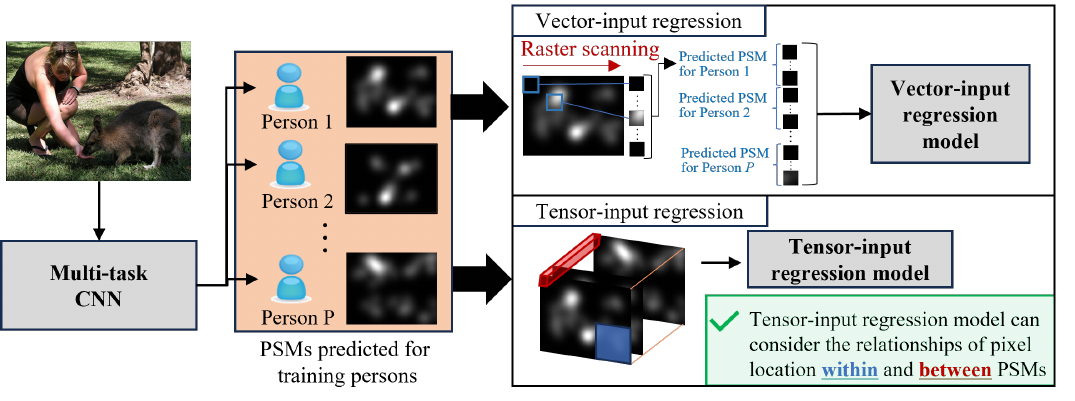}
    \caption{\small \textcolor{\ColorResponseAB}{Comparison of the tensor-input regression with the vector-input regression in the proposed approach. The tensor-input regression is superior to the vector-input regression in the points that it can preserve the structural information, which means the two-dimensional spatial configuration in this paper, within and between PSMs.}}
    \label{reg_comparison}
\end{figure*}

\subsection{PSM Prediction via Tensor-to-Matrix Regression}
\label{Regression}
This subsection presents the tensor-to-matrix regression model for the few-shot PSM prediction. \textcolor{\ColorResponseAB}{The comparison of the tensor-input regression with the vector-input regression is shown in Fig.~\ref{reg_comparison}. The tensor-input regression is superior to the vector-input regression in the points that it can preserve the structural information, which means the two-dimensional spatial configuration in this paper, within and between PSMs.}
\par 
\textcolor{\ColorResponseAB}{
In the proposed method, we do not explicitly define or compute gaze pattern similarity between individuals. Instead, the similarity is implicitly captured through the training process of the tensor-to-matrix regression model. Specifically, the model is trained to predict the target person's PSMs using the PSMs of training persons as input. During training, the weights are adjusted such that individuals with more similar gaze tendencies contribute more to the prediction. Therefore, similarity is learned as part of the regression process, without measuring similarities or prior clustering of individuals.}
\par The PSMs predicted in Sec.~\ref{multi-task-CNN} are used to predict the PSM of the target person. Several PSMs are treated as input. The input tensor $ \mathcal{S}(\bm{X}_{i}) \in \mathbb{R}^{P \times d_1 \times d_2}$ ($i=1,2,\dots,I$) corresponding to the image $\bm{X}_{i}$ chosen in Sec.~\ref{AIS} is constructed as follows:
\begin{align}
\mathcal{S}(\bm{X}_{i})= [\hat{\bm{S}}(\bm{X}_{i})_1,\hat{\bm{S}}(\bm{X}_{i})_2,\dots,\hat{\bm{S}}(\bm{X}_{i})_P].
\end{align}
In addition, we prepare the supervised PSM $\bm{S}(\bm{X}_{i})_{p^{\textrm{tst}}}$ of the target person $p^{\textrm{tst}}$ for the input tensor $ \mathcal{S}(\bm{X}_{i})$. Here, we assume that the target person gazes only at the common images selected in Sec.~\ref{AIS}, and we can obtain the supervised PSM $\bm{S}(\bm{X}_{i})_{p^{\textrm{tst}}}$. In a tensor-to-matrix regression scenario, the weight tensor $\mathcal{W} \in \mathbb{R}^{P \times d_1 \times d_2 \times d_1 \times d_2}$ is used to predict the PSM of a newly given image as follows:
\begin{align}
    \bm{S}_{\textrm{TReg}}(\bm{X}_{\textrm{tst}})_{p^{\textrm{tst}}}=\langle\mathcal{S}(\bm{X}_{\textrm{tst}}),\mathcal{W}\rangle_3,
\end{align}
where $\langle\cdot,\cdot\rangle_Q$ denotes the tensor product and $Q$ denotes the number of input arrays.
\par To optimize the weight tensor $\mathcal{W}$, we minimize the sum of the squared errors using $L_2$ regularization as follows:
\begin{align}
    \min_{\textrm{rank}(\mathcal{W})\leq R}\sum_{i=1}^I||\bm{S}(\bm{X}_{i})_{p^{\textrm{tst}}}-\langle\mathcal{S}(\bm{X}_{i}),\mathcal{W}\rangle_3||_F^2 + \lambda||\mathcal{W}||_F^2. \label{approximation_T2T}
\end{align}
Note that it is difficult to solve this minimization problem because the inputs and outputs are in a multi-array format. Thus, by referring to a previous study~\cite{lock2018tensor}, we assume that $\mathcal{W}$ has the reduced PARAFAC/CANDECOMP (CP)-rank such that $\textrm{rank}(\mathcal{W})\leq R$ and solve Eq. (\ref{approximation_T2T}) under this constraint. Although tensor-based regression tends to suffer from high-dimensional problems and overfitting, the low-rank approximation mitigates such problems. In addition, $L_2$ regularization suppresses overfitting. \textcolor{\ColorResponseAB}{The reduced CP-rank constraint serves as a dimensionality reduction strategy, which utilizes a multi-way structure to improve generalization and prediction efficiency. In this formulation, the ridge (L2) regularization is further interpreted as a Bayesian prior, which provides statistical motivation for the prediction procedure~\cite{lock2018tensor}.}
Therefore, using the tensor-to-matrix regression model, the proposed method preserves structural information without vectorizing the input tensor and output matrix.

\section{Experiments} 
\label{exp}
\subsection{Dataset}
\label{set}
In this experiment, an open-source large-scale PSM dataset~\cite{xu2018personalized} was used.
The PSM dataset was constructed such that the included images have high diversity and is suitable for performance and robustness evaluation.
The PSM dataset comprises 1,600 images with corresponding eye-tracking data obtained from 30 participants. 
The participants had normal or corrected visual acuity and gazed at one image for three seconds under free viewing conditions. 
To evaluate the predicted PSMs, we constructed the PSMs of each participant for all images from the eye-tracking data as the ground truth (GT) map based on a previous work~\cite{judd2009learning}. As the USM used in the proposed method, we adopted the mean PSMs of the training persons to reduce the influence of USM prediction errors.
\par 
In the proposed method, we required training images with eye-tracking data to train the multi-task CNN model and common images selected from the training images to train the tensor-to-matrix regression model. Thus, 1,100 images were randomly selected for training and the remaining 500 images were used as test images in the experiment. In addition, $I$ common images were selected from the training images based on the AIS scheme. \textcolor{\ColorResponseAB}{Note that the common images selected by AIS are used exclusively for training the prediction model. The evaluation is performed on a separate test set that is not influenced by the AIS for ensuring that the prediction results are not biased by the image selection mechanism.} In addition, we randomly selected 20 participants as training persons, and the remaining 10 persons were treated as target persons. \textcolor{\ColorResponseAB}{The number of training subjects and common images were determined empirically to achieve a practical balance between training and test data. The 20 training persons were randomly selected to ensure sufficient variability in gaze patterns. The $I$ common images were selected from training images via the AIS scheme. If the number of common images is too small, the prediction may become unstable due to depending on a limited set of gaze patterns. While, if the number of common images is too large, the target person should gaze at the massive number of images, which is unrealistic, although the prediction may become stable. In this way, we experimentally determined the number of common images, and its value is described in Sec.~\ref{exp-setting}.} Although the eye-tracking data of the target persons were available, we only used the eye-tracking data of the target persons as the common images for PSM prediction because we assumed that the target persons gazed at the common images. 
\subsection{Experimental Settings}
\label{exp-setting}
\par 
We optimized the multi-task CNN model in Sec.~\ref{multi-task-CNN} and the tensor-to-matrix regression model in Sec.~\ref{Regression}. The multi-task CNN model was optimized via the stochastic gradient descent~\cite{bottou2010large} by referring to a previous study~\cite{xu2018personalized}, and the number of layers ($L$), momentum, batch size, epoch, and learning rate were set to 3, 0.9, 9, 1,000, and 3.0$\times 10^{-5}$, respectively. In addition, the tensor-to-matrix regression model was optimized by simply differentiating the weight parameters with tensor unfolding. \textcolor{\ColorResponseAA}{Furthermore, we set $I=100$ and conducted additional experiments focusing on hyperparameter analysis by varying the hyperparameters of the tensor-to-matrix regression model over $R \in \{5,10,\dots,50\}$ and $\lambda \in \{0.01,0.1,\dots,10000\}$ to examine their effect on performance.}
\par To objectively evaluate the proposed method, we adopted several USM and PSM prediction methods as comparison methods. We adopted the following USM prediction methods: Signature~\cite{hou2012image}, GBVS~\cite{harel2007graph}, Itti~\cite{itti1998model}, SalGAN~\cite{pan2017salgan}, and Contextual~\cite{kroner2020contextual}. Signature, GBVS, and Itti are computational models that predict USMs only from input images. SalGAN and Contextual are deep learning-based models trained on the SALICON dataset~\cite{jiang2015salicon}, which is a large-scale eye-tracking dataset, without considering personalization. 
The following two few-shot PSM prediction (FPSP) methods using only common images and their eye-tracking data were adopted. 
\\
\textbf{Baseline1}: PSM prediction using visual similarity between target and common images~\cite{moroto2018gcce}.\\
\textbf{Baseline2}: PSM prediction based on Baseline1 and USM prediction~\cite{moroto2019ICCETW}.\\
In addition, we compared three PSM prediction methods with settings similar to those of the proposed method: similarity-based FPSP~\cite{moroto2020few}, CoMOGP-based FPSP~\cite{moroto2021few}, and object-based gaze similarity (OGS)-based FPSP~\cite{moroto2022few}. 
Note that although other PSM prediction methods exist~\cite{xu2017beyond,xu2018personalized,sherkati2022clustered,strohm2024learning}, they cannot learn from a small amount of training data \textcolor{\ColorResponseAA}{as discussed in Sec.~\ref{psm_prediction_related_works}}. Therefore, we adopted only the aforementioned comparison methods in our experiment.
\par 
As the evaluation metrics, we adopted the Kullback–Leibler divergence (KLdiv) and cross-correlation (CC) between the predicted PSM and the GT map based on previous research~\cite{bylinskii2018different}. Specifically, KLdiv was used to evaluate the similarity of the distribution, that is, structural similarity, and CC was used to evaluate pixel-based similarity. By using these two metrics, both the global and local similarities between the predicted PSMs and their GT maps can be evaluated.
\begin{figure*}[t]
  \begin{center}
    \includegraphics[clip,width=\textwidth]{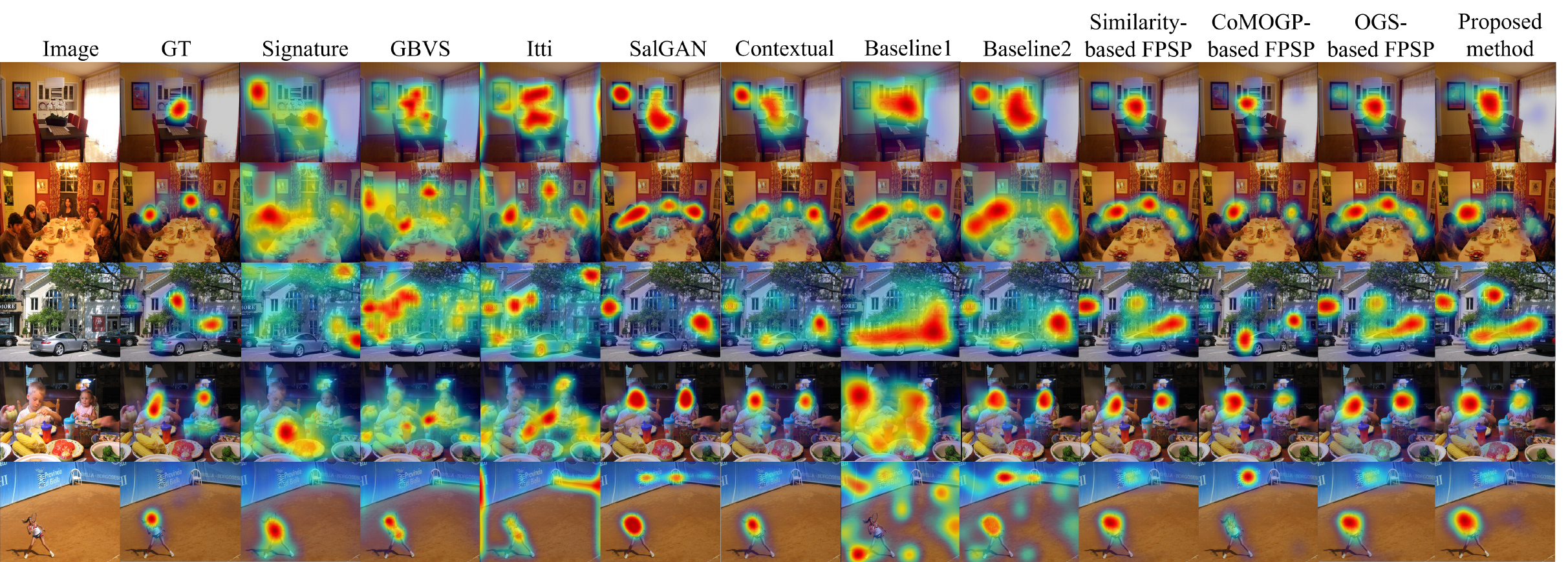}
    \caption{\small Examples of predicted PSMs.}
    \label{experiment}
  \end{center} 
\end{figure*}
\begin{table}[t]
  \caption{\small Quantitative distribution-based evaluation based on KLdiv and pixel-based evaluation based on CC. A lower KLdiv value indicates higher performance, whereas a higher CC value indicates higher performance.}
\centering
\small
\begin{tabular}{l|ccc}
          \hline
          \textbf{Methods} & \textbf{KLdiv}$\downarrow$ & \textbf{CC}$\uparrow$ \\ 
          \hline
          Signature~\cite{hou2012image}  &  8.04 & 0.413   \\ 
          GBVS~\cite{harel2007graph}     & 6.89 & 0.437    \\
          Itti~\cite{itti1998model}      & 9.04  & 0.322       \\
          SalGAN~\cite{pan2017salgan}   & 3.56 & 0.635      \\ 
          Contextual~\cite{kroner2020contextual}  & 3.57 & 0.674\\
          \hline
          Baseline1~\cite{moroto2018gcce}  & 7.64 & 0.401         \\ 
          Baseline2~\cite{moroto2019ICCETW}  & 4.13 & 0.597  \\
          Similarity-based FPSP~\cite{moroto2020few}  & 1.82  & 0.735  \\
          CoMOGP-based FPSP~\cite{moroto2021few} &1.38 & 0.765\\
          OGS-based FPSP~\cite{moroto2022few} & 1.09 & \textbf{0.781}\\         
          \hline
          Proposed Method ($R=50$, $\lambda=1000$) & \textbf{1.00} & 0.775\\
          \hline  
    \end{tabular}
    \label{outcome_table}
\end{table}
\begin{figure*}[t]
  \begin{minipage}[b]{0.48\linewidth}
    \centering
    \includegraphics[keepaspectratio, scale=0.50]{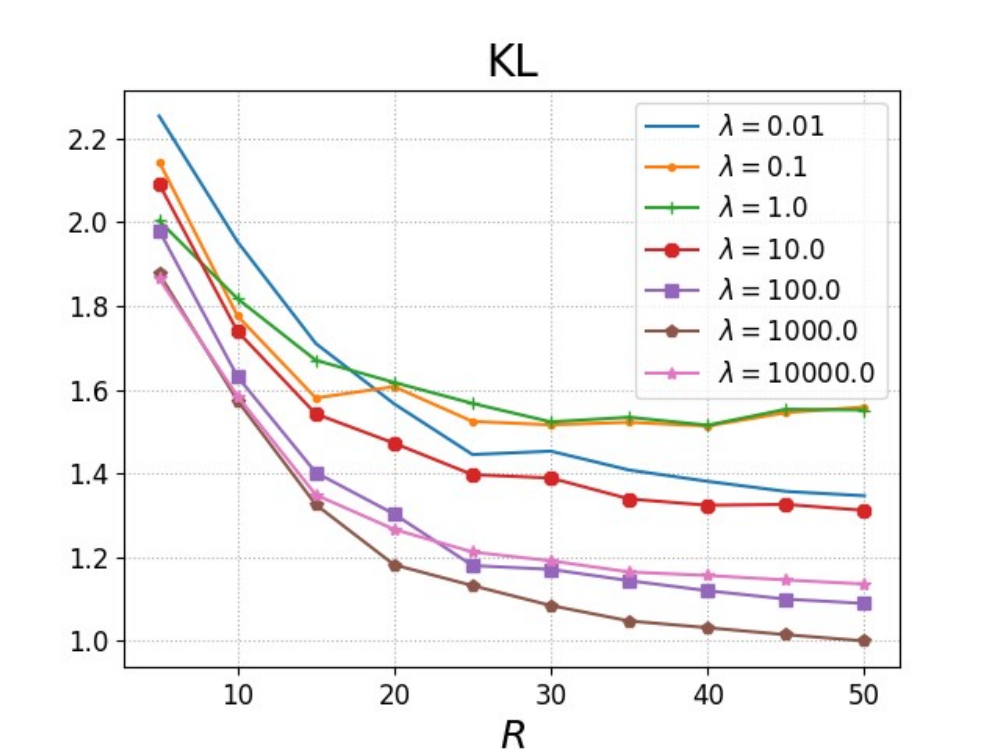}
  \end{minipage}
  \hspace{0.2mm}
  \begin{minipage}[b]{0.48\linewidth}
    \centering
    \includegraphics[keepaspectratio, scale=0.50]{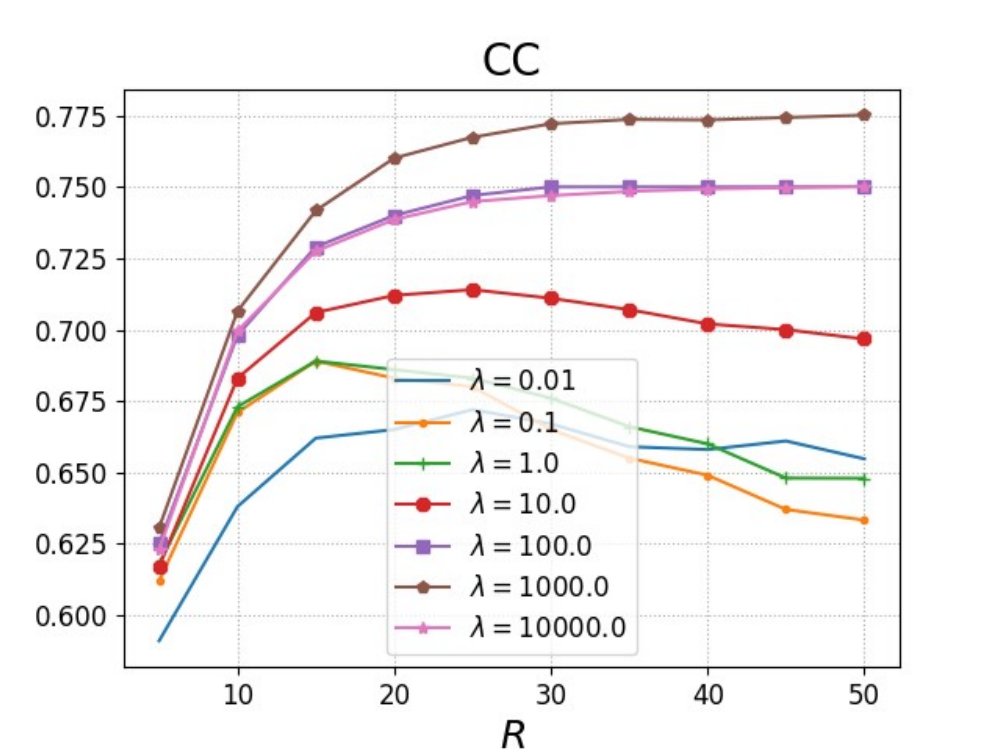}
  \end{minipage}
  \caption{\small Changes in the values of the evaluation metrics in response to changes in hyperparameters of tensor-to-matrix regression.}
  \label{hyperparameters}
\end{figure*}
\subsection{Results and Discussion}
\label{discussion}
Figure~\ref{experiment} presents the predicted results. Table~\ref{outcome_table} presents the quantitative evaluation results. As shown in Fig.~\ref{experiment}, the PSMs predicted by the proposed method exhibit a distribution close to the GTs, demonstrating the effectiveness of preserving structural information. In addition, as shown in Table~\ref{outcome_table} that compares the proposed and comparison methods, the proposed method outperforms all comparison methods in terms of KLdiv. This result confirms that tensor-to-matrix regression is effective for PSM prediction considering structural information. 
The effectiveness of personalization for saliency prediction was confirmed because the proposed method outperformed the USM prediction methods. In addition, by comparing the proposed method with other PSM prediction methods, the effectiveness of focusing on structural information is confirmed. Moreover, Baselines1 and 2 do not use the PSMs of other persons, and the comparison between these methods and the proposed method demonstrates the effectiveness of using eye-tracking data obtained from other persons. 
\par
The ``KLdiv'' of OGS-based FPSP~\cite{moroto2022few} is similar to that of the proposed method. 
\textcolor{\ColorResponseAA}{Here, OGS-based FPSP predicts PSMs by searching for objects in input images from the training images and using the PSMs corresponding to the regions of searched objects. In cases where an input image contains objects that were not present in the training dataset, OGS-based FPSP may be difficult to predict PSMs because it relies on PSMs of known objects. Consequently, its applicability may be limited when input images contain objects that were not seen during training. In contrast, the proposed method focuses on the relationships between persons' gaze patterns without semantic information, e.g., objects. Thus, our approach remains applicable even when the input images include objects that were not present in the training dataset. This characteristic allows the proposed method to be used in a broader range of scenarios.}
\par
\textcolor{\ColorResponseAA}{The ``CC'' of the proposed method is comparable to that of other PSM methods such as the OGS-based FPSP; however, it is not the best. One possible reason is that the expressive capacity of the model is limited by the rank $R$ in the low-rank tensor approximation of the weight tensor in Sec.~\ref{Regression}. In general, increasing $R$ enhances the model's ability to capture complex patterns, including fine-grained local information; however, it also substantially increases computational cost. The discussion of hyperparameters in the tensor-to-matrix regression model is described below.} \textcolor{\ColorResponseAB}{Another reason why the ``CC'' of the proposed method is lower than that of OGS-based FPSP is that the proposed method does not incorporate explicit semantic information such as object categories. In contrast, the OGS-based FPSP utilizes object-level matching to transfer saliency information, which results in higher CC scores. Additionally, the images used in our experiments contain a wide variety of general-purpose images, and many objects in the test images also appear in the training images. This enables the OGS-based FPSP to utilize its object-based mechanism effectively. However, as discussed above, the OGS-based FPSP has limited applicability when the test images include unseen objects. In such scenarios, its object-dependent mechanism cannot be applied. In contrast, the proposed method relies on interpersonal gaze similarity rather than object-specific information, which supports the applicability of the proposed method, particularly in real-world settings where unseen objects appear.}
\begin{figure*}[t]
  \begin{center}
    \includegraphics[clip,width=0.9\textwidth]{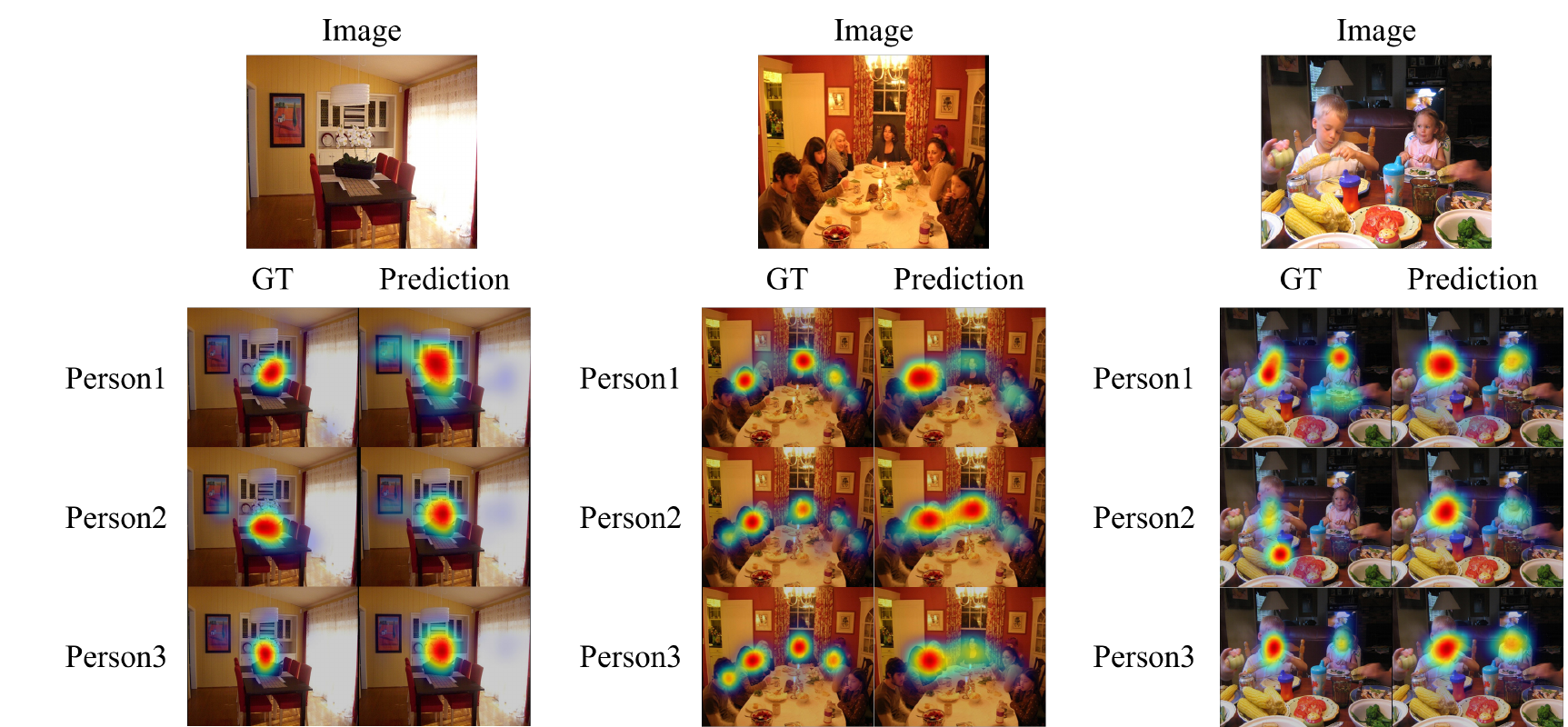}
    \caption{\small \textcolor{\ColorResponseAC}{Examples of PSMs predicted for several persons using the proposed method.}}
    \label{experiment2}
  \end{center} 
\end{figure*}
\par 
Here, ``CC'' is a pixel-based evaluation, whereas ``KLdiv'' is a distribution-based evaluation. Thus, the proposed method can preserve structural information because of its high ``KLdiv.'' Here, as mentioned in Sec.~\ref{intro}, structural information is an important clue to detecting salient areas in the human visual system~\cite{itti1998model}. From this viewpoint, we confirmed that the proposed method is valid because it is not significantly inferior to OGS-based FPSP in terms of ``CC'' and superior in terms of ``KLdiv.''
Therefore, we emphasize the effectiveness of the proposed method for PSM prediction by preserving the structural information. 
\par We confirmed the evaluation scores in response to changes in the hyperparameters of the tensor-to-matrix regression model through \textcolor{\ColorResponseAA}{additional experiments focusing on hyperparameter analysis}. Figure~\ref{hyperparameters} shows the evaluation scores in response to $R$ and $\lambda$. As shown in the figure, as $R$ increases, the performance improves, whereas $\lambda=1000$ achieves the best performance regardless of $R$. \textcolor{\ColorResponseAB}{Although these results were obtained experimentally, the trend was stable. Beyond this value, further increases in $\lambda$ led to performance degradation. Therefore, we adopted $\lambda=1000$ as a setting corresponding to a local maximum in performance.}
\textcolor{\ColorResponseAA}{The results also indicate that the performance improves as $R$ increases, but the improvement saturates around  $R = 50$. Beyond this point, further increases in $R$ lead primarily to longer computation times without meaningful gains in prediction performance. Therefore, we adopted $R = 50$ as a practical compromise between model performance and efficiency.}
In addition, an extremely high $\lambda$ value is not required because $\lambda$ is a regularization hyperparameter. Therefore, we confirm the desirable hyperparameters of tensor-to-matrix regression in the proposed method. \textcolor{\ColorResponseAA}{While this study limits the exploration to $R$, further investigation with more efficient model designs to reduce computational cost is a promising direction for future work.} \textcolor{\ColorResponseAB}{In addition, we adopted the CP-rank decomposition for the weight tensor in the tensor-to-matrix regression model by referring to a previous study~\cite{lock2018tensor}. The validation of other rank decomposition methods is also a future work.}
\par 
\textcolor{\ColorResponseAC}{
To show the difference in PSMs across individuals, we present the examples of PSMs predicted for several persons using the proposed method in Fig.~\ref{experiment2}. Most of these examples were correctly predicted by the proposed method. While some predicted PSMs in Fig.~\ref{experiment2} did not perfectly align with the actual gaze locations. For example, the main focus might be slightly shifted for certain individuals, such as Person2 in the right column. However, the proposed method still captured the overall attended regions with reasonable performance. It is important to note that human gaze behavior can vary not only due to the visual stimulus but also based on transient factors such as short-term intentions or long-term mental states. Therefore, perfectly predicting individual attention is inherently difficult. Despite this variability, the proposed method was able to approximate the attention tendencies of each person, which demonstrates its effectiveness in modeling personalized saliency.}
\section{Conclusions}
This study has proposed a few-shot PSM prediction method based on \textcolor{\ColorResponseAB}{interpersonal} gaze patterns. 
The proposed method incorporates the adaptive image selection scheme and tensor-to-matrix regression for effective image selection of images and the preservation of structural information, respectively.
By treating the input and output PSMs without vectorization, the proposed method preserves structural information. 
The experiments on the open dataset demonstrate the effectiveness of incorporating these factors.

\section*{Acknowledgement}

This work was partly supported by the JSPS KAKENHI Grant Numbers JP24K02942, JP23K21676, and JP23K11211.

\begin{biography}
\profile{Yuya}[]{Moroto}[]{
received his B.S. degree in Electronics and Information Engineering from Hokkaido University, Japan in 2019, and his M.S. and Ph.D. degrees in Information Science and Technology from Hokkaido University, Japan in 2021 and 2024. 
His research interests include computer vision, biological information analysis, multimodal signal processing and its applications. He is a member of IEEE.}

\profile{Keisuke}[]{Maeda}[]{
received his B.S., M.S., and Ph.D. degrees in Electronics and Information Engineering from Hokkaido University, Japan in 2015, 2017, and 2019. At present, he is currently a Specially Appointed Associate Professor in the Data-Driven Interdisciplinary Research Emergence Department, Hokkaido University. His research interests include multimodal signal processing and machine learning and its applications. He is an IEEE and IEICE member.
}

\profile{Takahiro}[]{Ogawa}[]{
received his B.S., M.S. and Ph.D. degrees in Electronics
and Information Engineering from Hokkaido University, Japan in 2003,
2005 and 2007, respectively.
He joined Graduate School of Information Science and Technology, Hokkaido University in 2008.
He is currently a professor in the Faculty of Information Science and Technology, Hokkaido
University.
His research interests are AI, IoT and big data analysis for multimedia signal processing and its applications.
He was a special session chair of IEEE ISCE2009, a Doctoral Symposium Chair of ACM ICMR2018, an organized session chair of IEEE GCCE2017-2019, a TPC Vice Chair of IEEE GCCE2018, a Conference Chair of IEEE GCCE2019, etc.
He has been also an Associate Editor of ITE Transactions on Media Technology and Applications.
He is a senior member of IEEE and a member of ACM, IEICE and ITE.
}

\profile{Miki}[]{Haseyama}[]{
received her B.S., M.S. and Ph.D. degrees in Electronics
from Hokkaido University, Japan in 1986, 1988 and 1993, respectively.
She joined the Graduate School of Information Science and Technology,
Hokkaido University as an associate professor in 1994. She was a visiting associate professor of Washington University, USA from 1995 to
1996. She is currently a professor in the Faculty of Information Science and Technology Division of Media and Network Technologies, Hokkaido University. Her research interests
include image and video processing and its development into semantic analysis.
She has been a Vice-President of the Institute of Image Information and Television Engineers, Japan (ITE), an Editor-in-Chief of ITE Transactions on Media Technology and Applications, a Director, International Coordination and Publicity of The Institute of Electronics, Information and Communication Engineers (IEICE).
She is a member of the IEEE, IEICE, Institute of Image Information and Television Engineers (ITE) and Acoustical Society of
Japan (ASJ).
}
\end{biography}
\end{document}